\documentclass[conference]{IEEEtran}
\usepackage[linesnumbered,ruled,vlined]{algorithm2e}
\usepackage{amssymb,amsmath}
\usepackage{caption}
\usepackage{subcaption}
\usepackage{comment}
\usepackage{color,soul}
\usepackage{ragged2e} 
\usepackage{algorithmicx}
\usepackage{algpseudocode}
\usepackage{amsthm}
\usepackage{cite}
\usepackage{array}

\usepackage{cite}

\ifCLASSINFOpdf
   \usepackage[pdftex]{graphicx}
   \graphicspath{{../}{../}}
   \DeclareGraphicsExtensions{.pdf,.jpeg,.png,.jpg}
\else
  \usepackage[dvips]{graphicx}
   \graphicspath{{../}}
   \DeclareGraphicsExtensions{.eps}
\fi

\usepackage[caption=false,font=footnotesize]{subfig}
\begin{document}

\title{An EV Charging Scheduling Mechanism to Maximize User Convenience and Cost Efficiency}

\author{\IEEEauthorblockN{
Hwei-Ming Chung\IEEEauthorrefmark{1}, Bahram Alinia\IEEEauthorrefmark{1}, No{\"e}l Crespi\IEEEauthorrefmark{1}, and Chao-Kai Wen\IEEEauthorrefmark{2}}

\IEEEauthorblockA{\IEEEauthorrefmark{1} {Institut Mines-T{\'e}l{\'e}com, T{\'e}l{\'e}com SudParis,
Evry, France}}
\IEEEauthorblockA{\IEEEauthorrefmark{2} Institute of Communications Engineering,\\
National Sun Yat-sen University, Kaohsiung, Taiwan 804} 

\IEEEauthorblockA{Emails: \IEEEauthorrefmark{1}\{hwei-ming.chung, bahram.alinia, noel.crespi\}@telecom-sudparis.eu, \IEEEauthorrefmark{2}chaokai.wen@mail.nsysu.edu.tw}

}

\maketitle
\begin{abstract}

This paper studies charging scheduling problem of electric vehicles (EVs) in the scale of a microgrid (e.g., a university or town) where a set of charging stations are controlled by a central aggregator. A bi-objective optimization problem is formulated to jointly optimize total charging cost and user convenience. Then, a close-to-optimal online scheduling algorithm is proposed as solution. The algorithm achieves optimal charging cost and is near optimal in terms of user convenience. Moreover, the proposed method applies an efficient load forecasting technique to obtain future load information. The algorithm is assessed through simulation and compared to the previous studies. The results reveal that our method not only improves previous alternative methods in terms of Pareto-optimal solution of the bi-objective optimization problem, but also provides a close approximation for the load forecasting. 
\end{abstract}

\section{Introduction}

\IEEEPARstart{I}{ssues} related to greenhouse gases emission by internal combustion engines in conventional vehicles have seen as the major factor accelerating the development of Electric Vehicles (EVs). 
Recent technology advancement in manufacturing efficient batteries  which are improving yearly by $20\%$ in terms of cost and $120 \%$ in terms of capacity while providing faster charging rate \cite{2014-EV-everywhere,2014-grid-of-future,IEA-report} makes EVs more attractive by .
Therefore, it is not surprising that the market witnessed $80\%$ increase in demand for EV \cite{sales}.

With huge population of EVs, efficient charging of their batteries becomes an important issue. A coordinated charging usually is preferred to uncoordinated mode which is known to have adverse impacts on the power grid such as increasing both peak load and total cost, and putting stress on distribution transformers \cite{2013-sha-EVinvest}. 
This paper considers a coordinated charging where there are a number of parking stations under control of a central aggregator.
The central aggregator is responsible for charging scheduling of EVs by controlling the charging rates as well as the starting and finishing time of the charging. The scheduling is done by gathering required information from EVs including demands, arrival times and deadlines as inputs of the scheduling algorithm. 

Two important criteria for evaluating efficiency of charging scheduling algorithms are total charging cost for aggregator and users' convenience level.
While there are many studies that address optimizing charging cost \cite{2012-He-EV-selection,2014-tang-EV-mincost,2016-shao-EV-mincost} and user convenience level \cite{2012-wen-EV-selection,2012-peter-EV-selection,2016-mal-EV-selection} separately, only a few studies consider both factors as the underlying merit factors \cite{2013-Ngu-EV-joint,2014-Tushar-EV-joint}.
Studies in \cite{2012-He-EV-selection,2014-tang-EV-mincost} try to minimize charging cost for the parking station owner where a similar research in \cite{2016-shao-EV-mincost} minimizes grid generation cost. 
Despite good results in \cite{2012-He-EV-selection,2014-tang-EV-mincost,2016-shao-EV-mincost}, users’ convenience level is ignored. 
The user convenience is considered in \cite{2012-wen-EV-selection,2012-peter-EV-selection,2016-mal-EV-selection} as the main objective of the scheduling problem to be maximized. 
In \cite{2012-wen-EV-selection}, user convenience is defined based on the total charging time of EVs and a distributed algorithm proposed to maximize it.  
\cite{2016-mal-EV-selection} considers the same problem while taking into account the power profile constraint imposed by the grid utility. 
Although user convenience is tried to be maximized in \cite{2012-wen-EV-selection,2012-peter-EV-selection,2016-mal-EV-selection}, the charging cost is not discussed and there is no cost effective strategy to optimize the cost for station or EV owners.
To address this issue, we consider joint optimization of user convenience and charging cost when designing scheduling algorithm. 
Study in \cite{2013-Ngu-EV-joint,2014-Tushar-EV-joint} have the same objective where in \cite{2013-Ngu-EV-joint} the authors propose a solution for minimizing total home electricity cost while considering user comfort level. 
However, the study in this paper is different than the study in \cite{2013-Ngu-EV-joint} since we focus on charging scheduling of EVs in multiple parking stations with charging rate limits and utilizing load forecasting.

When the future load in a charging station is not known, a forecasting method can be applied to improve the scheduling performance by not exceeding the available load at each time instant. 
For this purpose, short term load forecasting methods such as \cite{2007-taylor-loadpred-EUdata,2013-taylor-loadpredic} are needed. We will utilize the load forecasting in this paper to enhance the performance and accuracy of our proposed scheduling algorithm. 

Based on the above discussions, in this paper, we consider both cost and user convenience level in the objective function of the optimal charging scheduling problem for EVs. 
Our definition of user convenience is based on the State of Charge (SOC) of EVs' battery and the total time of charge. 
Since there is an inherent trade-off between these two factors, both objectives cannot be optimized simultaneously. 
Therefore, a solution which is good enough in terms of both objectives will be explored. 

The main contributions of this paper are as follows:
\begin{itemize}
\item We formulate a bi-objective optimization problem in a microgrid scale consisting of a set of local aggregators controlled by a central aggregator. Our study is inspired by \cite{2012-He-EV-selection} for charging cost minimization and \cite{2012-wen-EV-selection} for user convenience maximization to take into account both user convenience level and charging cost as the main performance factors in charging scheduling of EVs.

\item We propose a centralized and efficient scheduling algorithm to obtain a near-optimal solution for the formulated problem. The scheduling algorithm applies a reliable load forecasting technique in order to avoid exceeding the available load in each time slot. The designed algorithm is optimal in terms of the cost and close-to-optimal in terms of the user convenience. Moreover, we design our scheduling algorithm wisely to improve average charging time without affecting the main objective value.

\item To evaluate the proposed method, we conduct extensive simulations and the results reveal that while the proposed method achieves the same result in \cite{2012-He-EV-selection} for charging cost, it improves the user convenience level. Also, we show that the applied forecasting method which is around $3.5\%$ more accurate than the simple method used in \cite{2012-He-EV-selection}. Furthermore, comparison with study in \cite{2012-wen-EV-selection} (user convenience maximization) shows that our method improves charging cost significantly for a relatively small decrease in user convenience.

\end{itemize}

The rest of the paper is organized as follows.
System model and the problem formulation are described in Section \ref{sec:system_model_problem_formulation}.
The scheduling algorithm to solve the problem is presented in Section \ref{sec:algorithm}.
We evaluate the proposed method in Section \ref{sec:simulation} and give conclusions in Section \ref{sec:conclusion}.

\section{System Model and Problem Formulation}\label{sec:system_model_problem_formulation}
In this section, we explain our system model and define the performance factors that we take into account in the problem formulation and algorithm design. 
\subsection{System Model}\label{subsec:system_model}

\begin{figure}
\begin{center}
\resizebox{2.3in}{!}{%
\includegraphics*{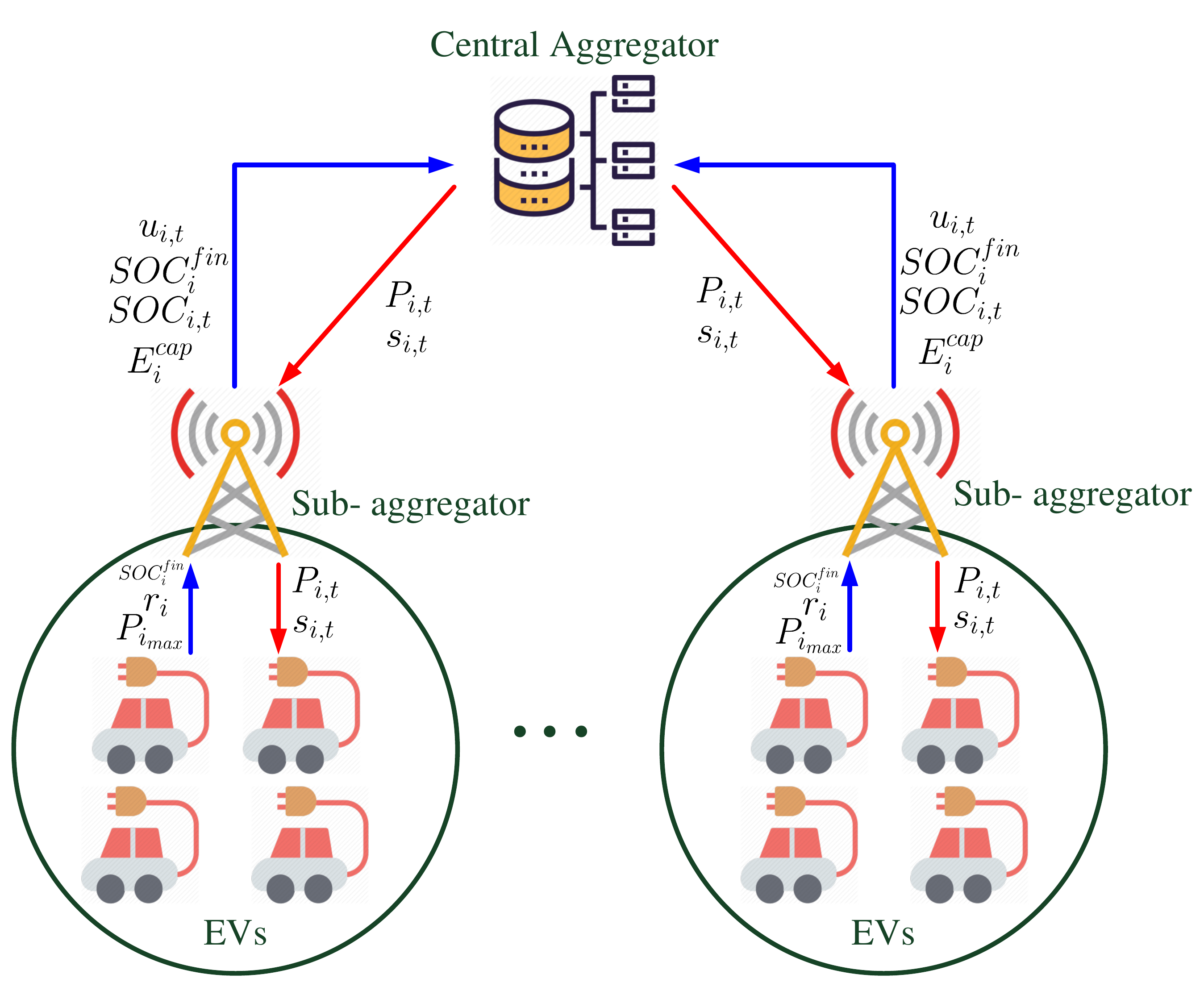} }%
\caption{The system model used in this paper. }\label{fig:cen_System_model}
\end{center}
\end{figure}

We study the charging scheduling problem of EVs in a microgrid (e.g., university or town) consisting of a set $\mathcal{B}$ of $M$ parking stations each of which having a sub-aggregator installed in the station and a central aggregator which controls all sub-aggregators to charge EVs' battery. This model is illustrated in Fig. \ref{fig:cen_System_model}.
The sub-aggregators receive final charging decisions through communication from the central aggregator.
The charging problem is studied in a time horizon of $T$ equally length time slots $t=1, 2,\dots ,T$  and total $N$ EVs in the charging stations.
EVs can recourse to the parking stations to charge their battery.
When an EV arrives, the sub-aggregator is able to catch all required information. These information will be sent to the central aggregator in order to provide an efficient scheduling of EVs in all parking stations. For an EV $i$, the State Of Charge (SOC) at time slot $t$ is denoted as $SOC_{i, t}$ with its range  between $0$ (for the empty battery) and $1$ (for the full battery). The target SOC denoted as $SOC_{i}^{fin}$ and indicates the SOC at the leaving time of EV $i$. $a_i$ and $r_{i}$ denote the arrival time and deadline, respectively. 
Note that the actual leaving time of an EV can be earlier than its deadline but cannot exceeds the user defined deadline. We use a $N \times T$  logical (or $0-1$) matrix $F$  to keep the track of available EVs in different time slots defined as follows: 
\begin{equation}\label{eq:charging_interval}
 f_{i, t} = \left\{
\begin{array}{lllcl}
1  ,   \mbox{EV i is in the station at time t} , \\
0,     \mbox{otherwise}.  
\end{array}
\right.
\end{equation}

Due to the physical constraints of batteries, maximum charging rate of EV $i$ is restricted to $P_{i_{max}}$ in each time slot $t$. 
Moreover, EV $i$'s battery capacity is denoted by $E_{i}^{cap}$. 
$E_{i}^{ini}$ denotes the initial SOC of EV $i$ in its arrival time i.e., $SOC_{i, a_i}E_{i}^{cap}$. According to the target SOC, the energy level of EV $i$ when it leaves the charging station should be greater than $SOC_{i}^{fin} E_{i}^{cap} $.

\begin{figure}
\begin{center}
\resizebox{2.5in}{!}{%
\includegraphics*{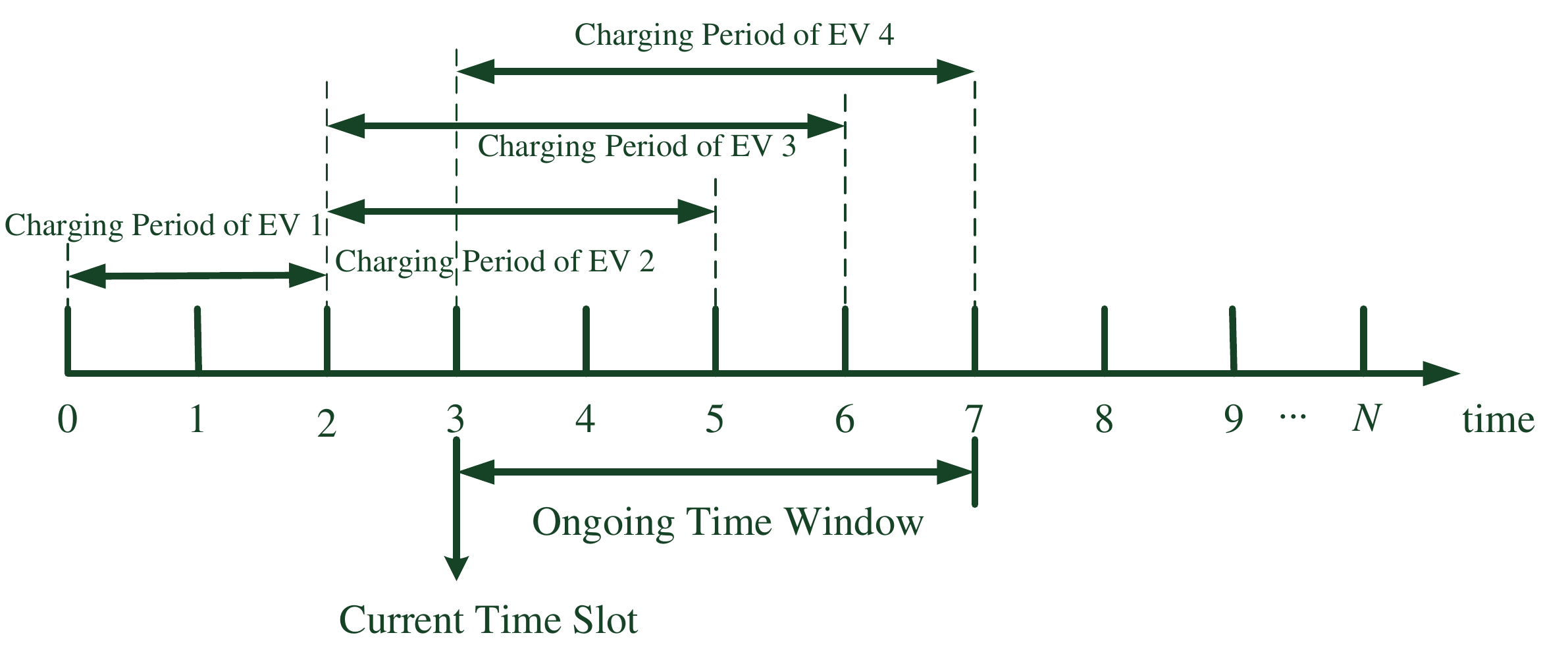} }%
\caption{Illustration of the current EV set and the ongoing time window\cite{2012-He-EV-selection}. }\label{fig:EV_time_set}
\end{center}
\end{figure}

The sub-aggregators has no information about the future coming EVs and the charging decisions need to be made at the beginning of each time slot based on the current available EVs. 
In each time slot $t$, $H_t$ denotes the current set of EVs which are available for charge at time time slot $t$. Moreover, $W_t$ is current sliding time window calculated at time slot $t$ which covers time slots $t$ to $t', t'>t$ where $t'$ is the latest deadline of EVs in set $H_t$.
Fig.\ref{fig:EV_time_set} illustrates the current EV set and sliding window at time slot 3. 
As shown in the figure, at the beginning of time slot 3, EV 1 already left the station and EVs 2, 3 and 4 are still in the charging station. 
Therefore, the current EV set in time slot 3, $H_3$, includes EVs 2, 3 and 4.
Furthermore, in the set $H_3$, EV 4 is the last one that leaves the station at time slot 7.
Hence, the current sliding window spans from time slot 3 to 7 i.e., $W_3= \{3, 4, 5, 6, 7\}$.

\subsection{Charging Cost}
The first performance factor that we consider in this paper is the electricity cost. 
In this regard, we borrow the following electricity price function from \cite{2012-He-EV-selection}:
\begin{equation}\label{eq:simple_cost}
	g(z_{t}) = k_{0} + k_{1} z_{t},
\end{equation}
The function $g(z_{t})$ returns the electricity cost for the total load $z_t$ at time slot $t$ where $K_0$ and $K_1$ are constants. 
Based on the price function, we can obtain the cost for EV charging part as follows
\begin{equation}\label{eq:int_cost}
	C_{t} = \int_{L_{base_{t}}}^{z_{t}} (k_{0} + k_{1} z_{t} )~~~d z_{t}
\end{equation}
where $L_{base_{t}}$ is the base load at time slot $t$. 
Finally, the second performance objective in this paper i.e., electricity cost for charging, $J_2$ can be shown as follows:
\begin{equation}\label{eq:obj2}
J_{1}=\sum_{t=1}^{T} \left( \left( k_{0}+\frac{k_{1}}{2} z_{t}^{2}\right) - \left( k_{0}L_{base_{t}}+\frac{k_{1}}{2} L_{base_{t}}^{2} \right) \right).
\end{equation}

\subsection{User Convenience}\label{subsec:weight}
In order to measure the users' satisfaction level, we define a parameter called user convenience based on a similar definition given by \cite{2012-wen-EV-selection}.
The user convenience level is an instant variable changing in each time slot during the stay of an EV in the charging station. 
For an EV $i$, we define the user convenience as the ratio of minimum possible charging time of EV $i$ and the remaining time to its deadline \cite{2012-wen-EV-selection}:
\begin{equation}\label{eq:weight}
u_{i, t} = \frac{w^{*}_{i, t}}{w_{i, t}} .
\end{equation}

In the Eq. (\ref{eq:weight}), $u_{i, t}$ is the user convenience level for EV $i$ at time slot $t$ which is defined based on two parameters $w^{*}_{i, t}$ and $w_{i, t}$. 
The first parameter, $w^{*}_{i, t}$, indicates the minimum required time slots to finish charging of EV $i$.
Putting it in another way, $w^{*}_{i, t}$ indicates that charging of EV $i$ cannot be finished earlier than time slot $t+w^{*}_{i, t}$ under any feasible scheduling. $w^{*}_{i, t}$ formally is defined as follows:
\begin{equation} \label{eq:minimum_waiting_time}
w^{*}_{i, t}   = \frac{ (SOC_{i}^{fin} - SOC_{i, t} ) E_{i}^{cap}}{P_{i_{max}}}. 
\end{equation}
The second parameter in the definition of user convenience level i.e., $w_{i, t}$, determines the number of remaining time slots to finish charging of EV $i$ before its deadline $r_i$:
\begin{equation} \label{eq:parking_time}
w_{i, t} = r_{i} - t +1.
\end{equation}

Based on the definition of the user convenience level in this section, the maximum user convenience can be obtained when in each time slot we charge the EV with its maximum charging rate. 
Therefore, if we sum up all user convenience level for an EV in different time slots during its charging, the obtained value is directly depended on the charging time of the EV.
Therefore, a good charging strategy may be maximizing the sum of all user conveniences for all EVs and all time slots, $J_2$, defined as follows:
\begin{equation}\label{eq:obj1}
J_{2}= \sum_{t =1}^{T} \sum_{i = 1}^{N} u_{i, t},
\end{equation}

The intuition behind the definition of user convenience is that at the beginning of each time slot we should give charging priority to EVs that are close to their deadline and received less electricity.

\subsection{Problem Formulation}\label{subsec:problem}
Now we can formulate the bi-objecitve optimization problem taking into account charging cost, $J_1$ and user convenience level, $J_2$ framed as follows:
\begin{subequations} \label{eq:total_for}
\renewcommand\minalignsep{0em}
\begin{align} 
&\mathcal{Z}: \min_{z, s , P, u }  quad quad \left\{J_{1},\frac{1}{J_{2}} \right\}&    \label{eq:total_obj}\\
&uaduad\mbox{subject to}    & \notag \\
&               uaduad z_{t} \leq L_{base_{t}} + \sum_{i=1}^N P_{i, t} ,        & \forall t, \label{eq:total_peak}\\
&		        uaduad E_{i}^{ini}+\sum_{t=1}^T P_{i, t} \leq E_{i}^{cap},       & \forall t, \label{eq:total_no_over}\\
&               uaduad P_{i, t}  \leq f_{i,t} P_{i_{max}} ,              & \forall t, i, \label{eq:total_power_cons}\\
&		        uaduad P_{i, t}\geq 0,   & \forall t,i, 
\label{eq:positive_charge}
\end{align}
\end{subequations}

The objective function in Eq. \ref{eq:total_obj} includes both cost and user convenience. 
Constraint Eq. \ref{eq:total_peak} states that the total load at time slot $t$ i.e., $z_t$ should be less than the summation of based load and total electricity flowed to the EVs at time slot $t$. Constraint Eq.\ref{eq:total_no_over} ensures that the total electricity in each EV's battery is less than or equal the battery capacity in the leaving time of the EV. Constraint Eq.\ref{eq:total_power_cons} takes care of charging rate limitation in EVs which should lie between 0 and maximum rates. When $P_{i,t}=0$, it means that EV $i$ is not selected to be charged at time slot $t$. 

\section{Solution}\label{sec:algorithm}

\subsection{Solution Approach}\label{sec:approach}
Joint minimization of both objectives of problem $\mathcal{Z}$ is not possible since there is an inherent trade-off between the electricity cost and user convenience. 
Therefore, in this section, we propose an algorithm to obtain a pareto optimal solution for the problem to achieve a solution which is good in terms of both objectives. 

The proposed algorithm works in two phases. 
In the first phase, we ignore user convenience ($J_2$) and find a solution which is optimal in terms of electricity cost ($J_1$). Put it in another way, we solve the following problem $\mathcal{P}_1$ in the first phase:
\begin{align}
\mathcal{P}_1: \ &\min_{} uad J_2 \label{P1} \\
&\ \textrm{s.t.}\ \ \ \textrm{constraints}\ \ref{eq:total_peak}-\ref{eq:positive_charge}\nonumber
\end{align}

By obtaining optimal solution of problem $\mathcal{P}_1$, we have a feasible solution for problem $\mathcal{Z}$ since the constraint sets are the same. Besides, this optimal scheduling of problem $\mathcal{P}_1$ gives the information on the optimal load values $z^\star_t$ for all time slots $t=1,\dots ,T$ to minimize the cost. Therefore, if we re-schedule EVs in the optimal solution of $\mathcal{P}_1$ by having the same optimal loads i.e., $z^\star_t, t=1,\dots ,T$ then, the new scheduling will be also an optimal solution for problem $\mathcal{P}_1$. 

In the second phase, we take $z^\star_t, t=1,\dots ,T$ from optimal solution of problem $\mathcal{P}_1$ with the optimal objective value $\Phi^\star$ and try to solve the following  problem $\mathcal{P}_2$: 

\begin{align}
\mathcal{P}_2 : \ &\min_{} uad \frac{1}{J_2} \label{P2} \\
&\ \textrm{s.t.}\ \ \ z_t=z^\star_t, t=1,\dots ,T\\
&\ uaduad \textrm{constraints}\ \ref{eq:total_no_over}, \ref{eq:total_power_cons}, \ref{eq:positive_charge}\nonumber
\end{align}

A feasible solution of $\mathcal{P}_2$ will be an optimal solution of $\mathcal{P}_1$. Then, inside the feasible solution of problem $\mathcal{P}_2$ we will try to maximize user convenience. 
Therefore, a close-to-optimal solution for $\mathcal{P}_2$ is an optimal solution for $\mathcal{P}_2$ and hence, a feasible and close-to-optimal solution for problem $\mathcal{Z}$. In the remaining of this section we explain the strategy to obtain a close-to-optimal solution for $\mathcal{P}_2$.

\subsection{Scheduling Algorithm}

Based on the given solution idea in Section \ref{sec:approach}, we present our scheduling algorithm in this section to solve problem $\mathcal{P}_2$. 
To tackle problem $\mathcal{P}_2$, we first solve problem $\mathcal{P}_1$ to obtain optimal charging cost.
Then, we try to maximize user convenience while keeping the optimal cost. To obtain optimal objective value of $\mathcal{P}_1$, we use presented algorithm in \cite{2012-He-EV-selection}. 

The core algorithm for solving problem $\mathcal{Z}$ is listed in Algorithm \ref{EV_sel}. 
The designed Charging Scheduling Algorithm (CSA) is inspired by the algorithms proposed in \cite{2012-He-EV-selection} for a minimum cost EV charging and \cite{2012-wen-EV-selection} for maximum user convenience. We stress that none of the algorithms in \cite{2012-He-EV-selection} and \cite{2012-wen-EV-selection} consider joint optimization of user convenience and charging cost. Moreover, for estimating the future load, we use a more effective load forecasting method described in Section \ref{subsec:load_forecasting}.
The CSA algorithm is intended to be run in central aggregator. Then, scheduling will be broadcast to all sub-aggregators. After solving $\mathcal{P}_1$, the available power at time slot $t$, $L_{t}$, can be obtained as follows:
\begin{equation}
z_{t}=z_t^\star - L_{base_{t}}
\end{equation}

By knowing the value of $z_t$ in each time slot $t$, Algorithm CSA calls sub-procedure User Convenience Maximization (UCM) described in Algorithm \ref{UCM} to schedule the EVs charging. The scheduling is done by applying Earliest Deadline First (EDF) policy on the EVs in the set $H_t$ and charging EVs with their maximum charging rate. More specifically, assume that EVs in set $H_t$ are sorted based on their deadline, $r_i$ in an increasing order. Then, UCM will go over the sorted list and select as many as possible number of EVs  to charge in time slot $t$ with their maximum charging rate (except probably the last selected EV) such that the total charging of selected EVs is equal to available load $z_t$. Note that this scheduling policy is feasible since the EDF is proved to be feasible. The strategy that UCM uses in charging of EVs helps to reduce average charging time of EVs and hence, improves user convenience level. When the charging decisions are made for time slot $t$, the SOC information needs to be updated as follows:
\begin{equation}\label{eq:SOC_update}
SOC_{i, t+1} = SOC_{i, t} + \frac{P_{i, t}}{E_{i}^{cap}}.
\end{equation}
If the $SOC_{i, t+1}$ reaches the target SOC then the charging task of EV $i$ is done and $t+1$ is considered as leaving time of the EV.

\begin{algorithm}\label{ago:EV_sel} 
\caption{EV Charging Scheduling Algorithm (CSA)}
\label{EV_sel}
 \DontPrintSemicolon 
\KwIn{$N$ EVs with their charging profile }
\KwOut{A feasible scheduling for problem $\mathcal{Z}$}
Forecast the base load  with load forecasting method in Section \ref{subsec:load_forecasting};

\For{t  = 1 \KwTo T }{
   Determine $H_t$ and $W_t$ based on the current time t\;
   Construct $F$ based on Eq.(\ref{eq:charging_interval})\;
   $E_{i}^{ini} \leftarrow SOC_{i, t} E_{i}^{cap}$ \;
   Solve problem $\mathcal{P}_1$ using interior point methods in \cite{Convex_opt} to obtain $z_{t}^\star$\;
   $L_{t} \leftarrow z_{t}^\star - L_{base_{t}}$\;   
   Call $UCM(L_t)$ to solve $\mathcal{P}_2$ with $H_t$ and $W_t$  \;  
    Use Eq.(\ref{eq:SOC_update}) to update SOC information \; 
   }
\end{algorithm}

\begin{algorithm}\label{ago:UCM} 
\caption{Algorithm $UCM(L_t)$}
\label{UCM}
 \DontPrintSemicolon 
\KwIn{Available power $L_{t}$, current time slot $t$ }
\KwOut{Charging decisions for time slot $t$}

Sort EVs in $H_t$ based on their deadline (EDF) as $e_1,e_2,\dots ,e_{|H_t|}$\;
$L_{\textrm{remain}}\leftarrow L_t$ \;

$i=1$\;

\While{$L_{\textrm{remain}}>0\ \wedge\ i\leq |H_t|$}{
$\Delta=\min \{L_{\textrm{remain}},P_{i_{max}}\}$

Charge $e_i$ with the rate of $\Delta$

$L_{\textrm{remain}}\leftarrow L_{\textrm{remain}}-\Delta$

$i\leftarrow i+1$}
\end{algorithm}

\subsection{Load Forecasting Method}\label{subsec:load_forecasting}
The load forecasting method mentioned in Algorithm \ref{EV_sel} is described in this section.
The seasonal autoregressive integrated moving average (ARIMA) model in \cite{2007-taylor-loadpred-EUdata} is employed as our forecasting technique where it is a generalization of the autoregressive moving average (ARMA) model first proposed by  George Box and Gwilym Jenkins and called Box-Jenkins model \cite{ARIMA-model}. Seasonal ARIMA is a well developed and stable algorithm which is used in this paper.
The advantage of seasonal ARIMA to ARMA is that if there is a correlation between data then the ARIMA model shows a better performance based on the the forecasting result.
Also, ARIMA applies different models of prediction. We refer to \cite{2007-taylor-loadpred-EUdata}  for further details. We calculate the forecasting error based on the Mean Absolute Percentage Error (MAPE) as follows:

\begin{equation} \label{eq:MAPE}
MAPE = \frac{1}{T} \sum_{i=1}^{T} |APE_{i}|. 
\end{equation}

The Average Percentage Error (APE) is defined as

\begin{equation} \label{eq:APE}
APE_{i} = \frac{L_{fore_{i}} - L_{act_{i}}}{max(L_{act})},
\end{equation}

where $L_{fore}$ and $L_{act}$ are the forecast and actual load, respectively.

\section{Simulation Results}\label{sec:simulation}
In this section, we evaluate our proposed scheduling algorithm in terms of load forecasting accuracy and objective value of problem $\mathcal{Z}$. Unless otherwise specified, the simulation setting is as follows: 

We consider a central aggregator which controls 4 local aggregators. Each local aggregator is installed in a parking station. The total number of EVs is 150 where they are randomly assigned to one of the 4 parking stations. 
The EVs used in the simulation are Nissan Leaf 2016 with battery capacity of 30 Kwh and maximum charging rate of 6.6 Kwh. In the pricing model in Eq.(\ref{eq:simple_cost}), the constants $K_{0}$ and $K_{1}$ are set to $10^{-4}$ C\$/Kwh and $1.2\times 10^{-4}$ C\$/Kwh/Kw, respectively as in \cite{2012-He-EV-selection}. The time horizon is divided to 96 time slots with the length 15 minutes to represent a 24 hours period. The arriving time and deadline of EVs are generated randomly between  $18:00$ and $07:00$, respectively. The feasible initial and final SOC values are randomly and uniformly generated from interval $[0,1]$.

In the simulation figures of this section, we compare three algorithms including CSA (proposed method in this paper), Method of \cite{2012-He-EV-selection} (cost minimization algorithm in \cite{2012-He-EV-selection}) and, Method of \cite{2012-wen-EV-selection} (user convenience maximization algorithm in \cite{2012-wen-EV-selection}). Since problem $\mathcal{Z}$ has two objective functions which cannot be optimized simultaneously, we define an inverse linear combination of the two objectives (i.e., $J_1$ and $J_2$) and compare different methods using this. In this regard, we define a merit factor $Q$ as follows:

\begin{equation}
Q=\frac{1}{\alpha \bar{J}_1+(1-\alpha) \frac{1}{\bar{J}_2}} 
\label{eq:Q}
\end{equation}

In Eq. (\ref{eq:Q}), $\bar{J}_1$ and $\bar{J}_2$ are normalized values of $J_1$ and $J_2$ respectively to lie in interval $[0, 1]$. We use the normalized values since $J_1$ and $J_2$ are in different scales. $\alpha$ is an application dependent parameter to weight charging cost and user convenience. With $\alpha =0$, the comparison is only based on user convenience and with $\alpha =1$ the effect of user convenience is ignored and three methods will be compared based on the charging cost. Based on the definition of $Q$, a higher value of $Q$ shows a better performance. To solve the optimization problem in \cite{2012-He-EV-selection} and $\mathcal{P}_1$ in this paper, we use CVX \cite{CVX}, a package intended to solve convex programs.

\subsection{Evaluating Forecasting Accuracy}
\begin{figure}
\begin{center}
\resizebox{1.8in}{!}{%
\includegraphics*{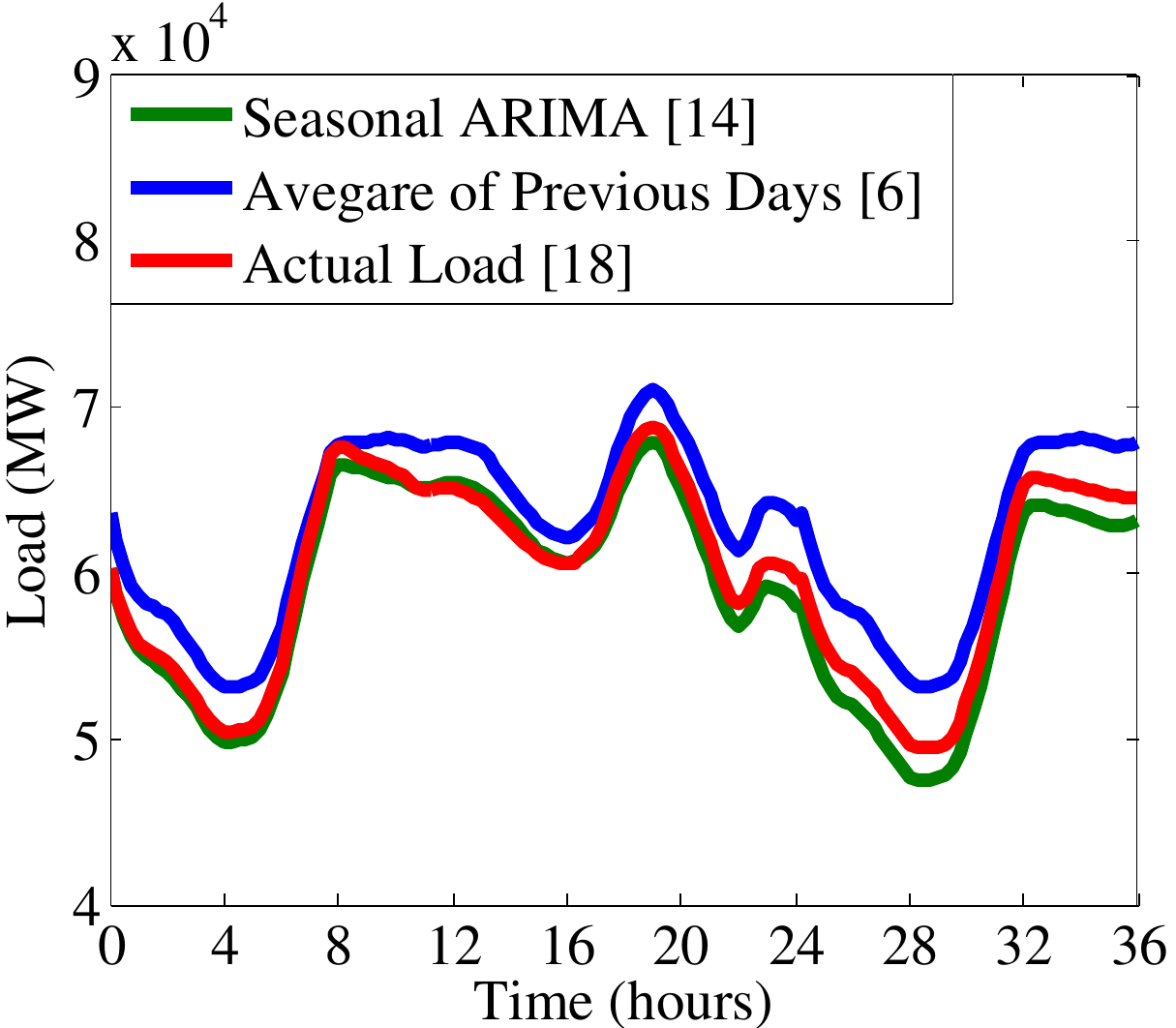} }%
\caption{Accuracy of forecasting methods.}\label{fig:load_1217_1218}
\end{center}
\end{figure}

In this section, we compare the forecasting method that we used in our scheduling algorithm (ARIMA method in \cite{2007-taylor-loadpred-EUdata}) with previous days' average method used in \cite{2012-He-EV-selection}. We used actual load information in France between 11/30/2015 to 12/16/2015 \cite{Load_data}.

As it can be seen from Figure \ref{fig:load_1217_1218}, the ARIMA method forecasts the future load with a high accuracy. The forecasting error for seasonal ARIMA and previous days' average method used in \cite{2012-He-EV-selection} are $2.67 \%$ and $6.77\%$, receptively which indicates a $4.1 \%$  improvement for ARIMA method. The other important point is that the forecast load by ARIMA is always below the actual load. This can improve system reliability and stability in practice. However, this is not true for previous days' average method.

\subsection{Evaluation Under Different Number of EVs }
\begin{figure*}[t!]	
	\centering
	\begin{subfigure}[b]{0.19\textwidth}
		\begin{center}
			\includegraphics[width=\textwidth]{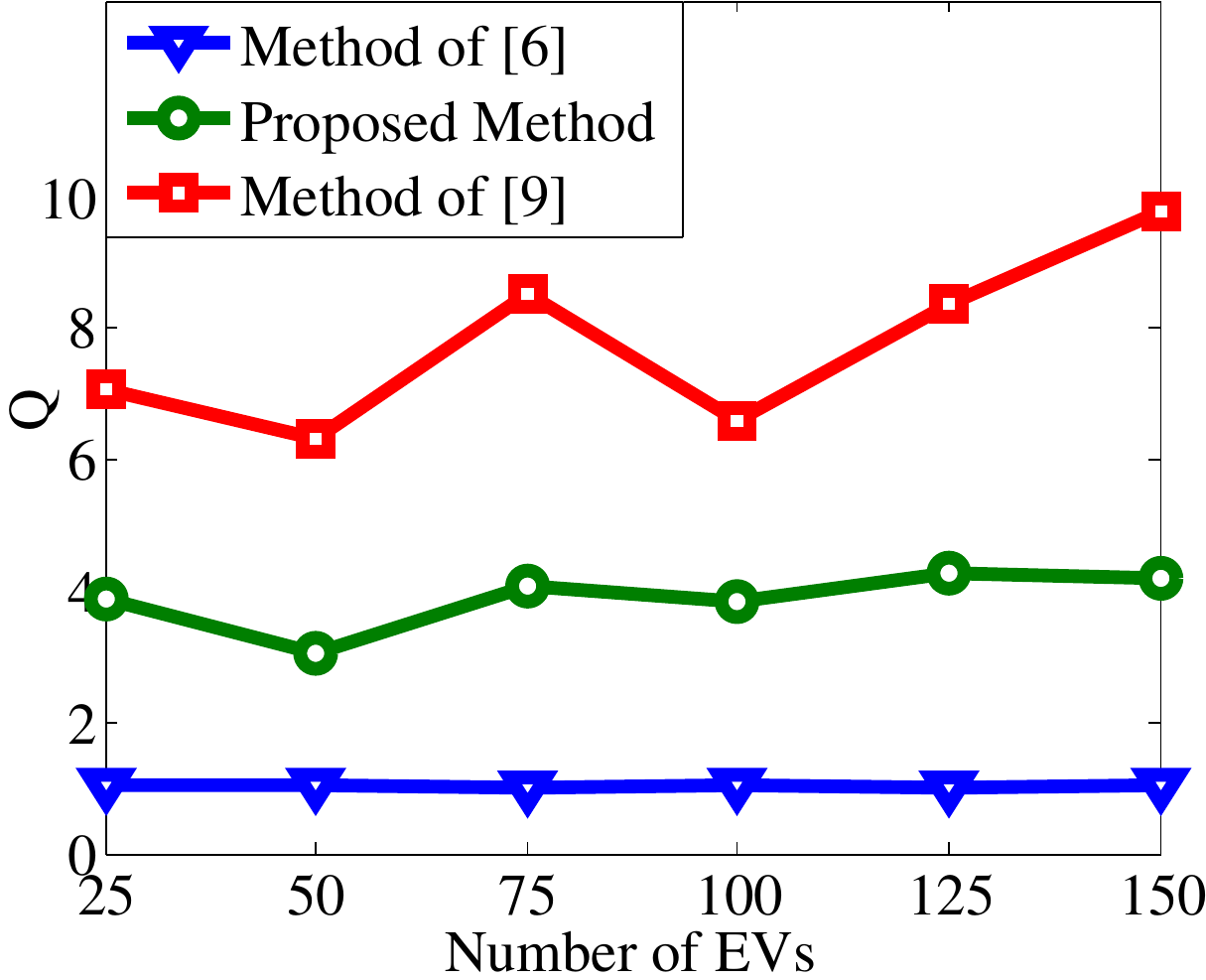}
			\caption{$\alpha=0$}
			\label{fig:alpha0}
		\end{center}
	\end{subfigure}%
	~ 
	\begin{subfigure}[b]{0.19\textwidth}
		\begin{center}
			\includegraphics[width=\textwidth]{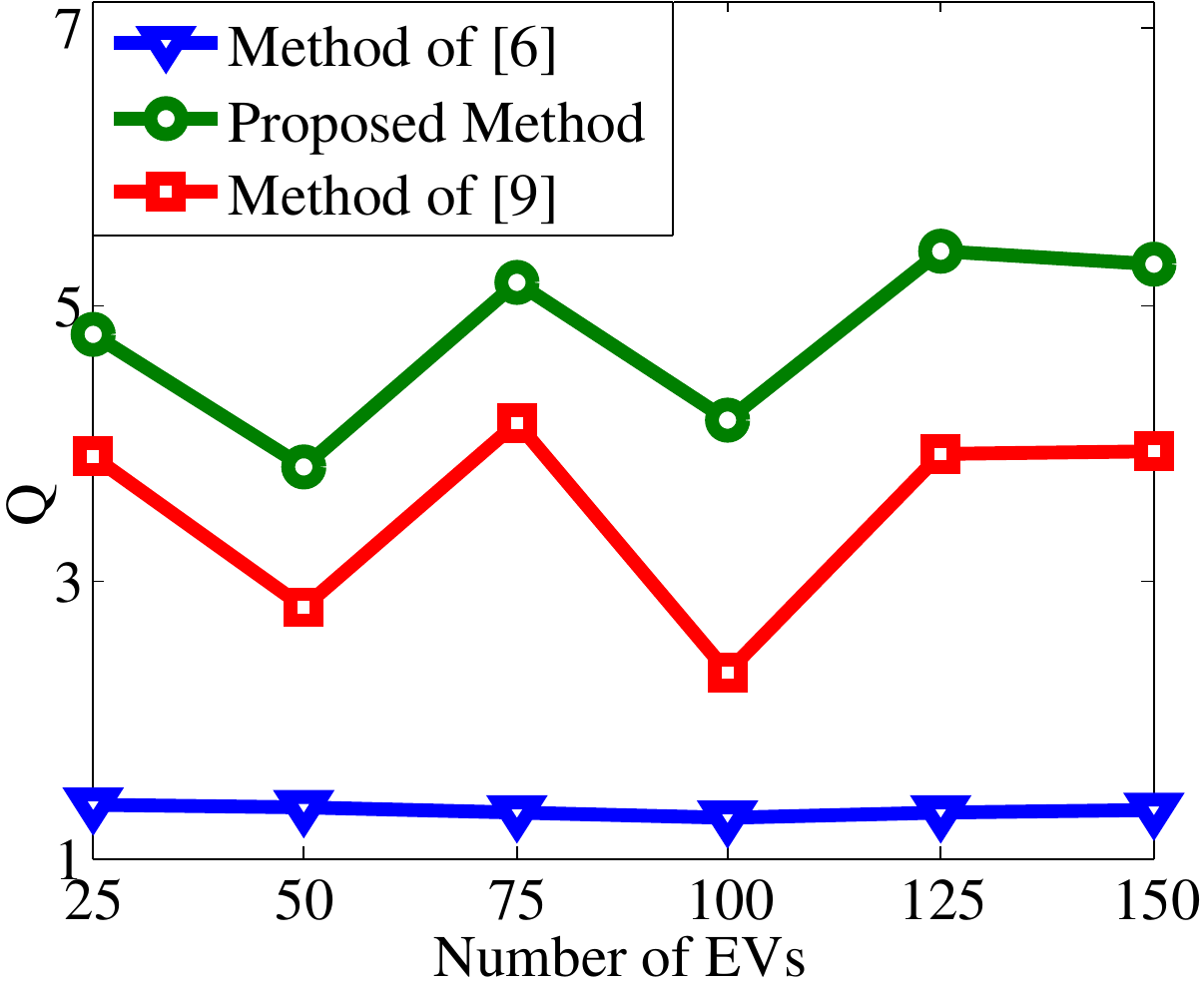}
			\caption{$\alpha=0.25$}
			\label{fig:alpha25}
		\end{center}
	\end{subfigure}
	\begin{subfigure}[b]{0.193\textwidth}
			\begin{center}
				\includegraphics[width=\textwidth]{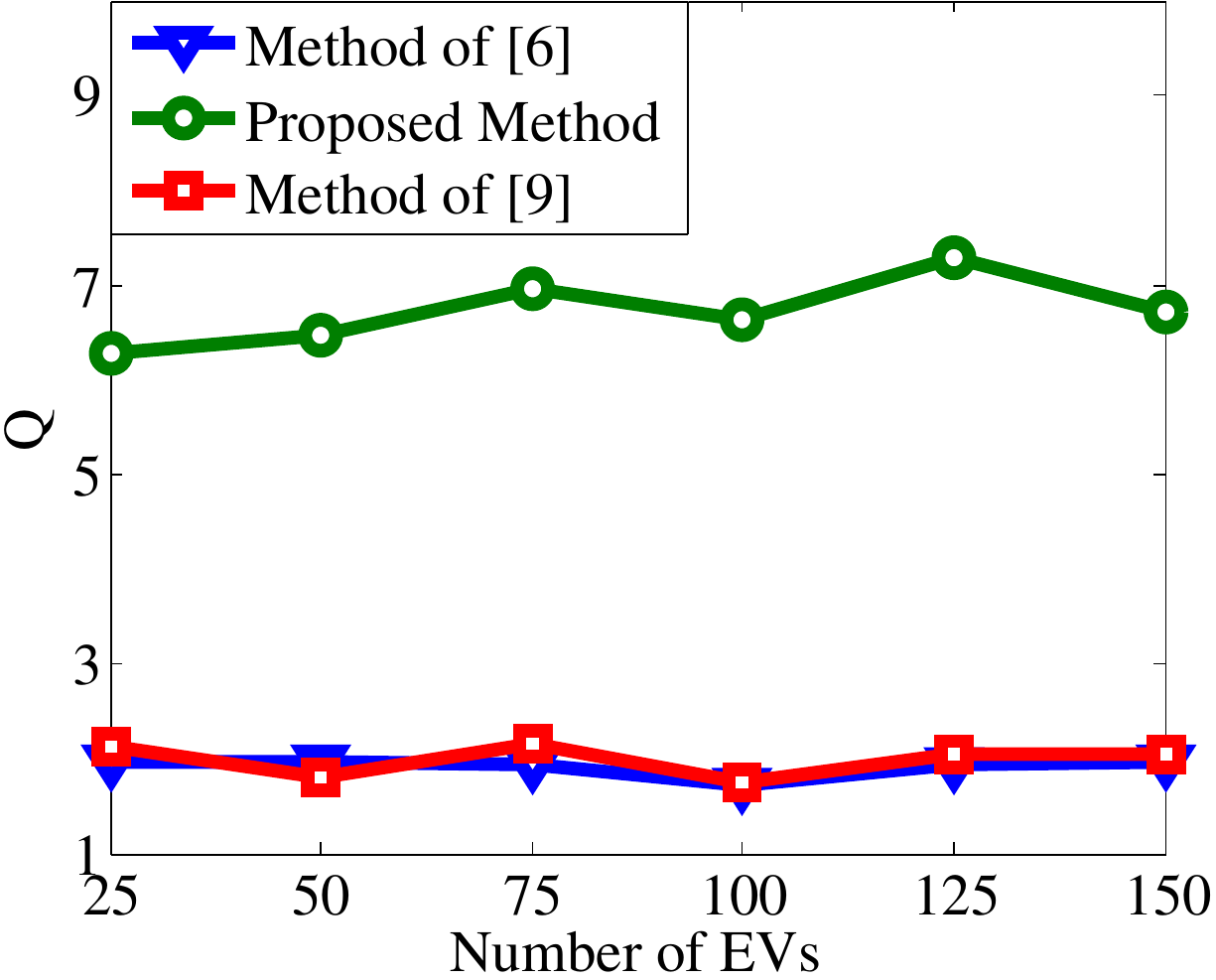}
				\caption{$\alpha=0.5$}
				\label{fig:alpha5}
			\end{center}
		\end{subfigure} 
		\begin{subfigure}[b]{0.192\textwidth}
			\begin{center}
				\includegraphics[width=\textwidth]{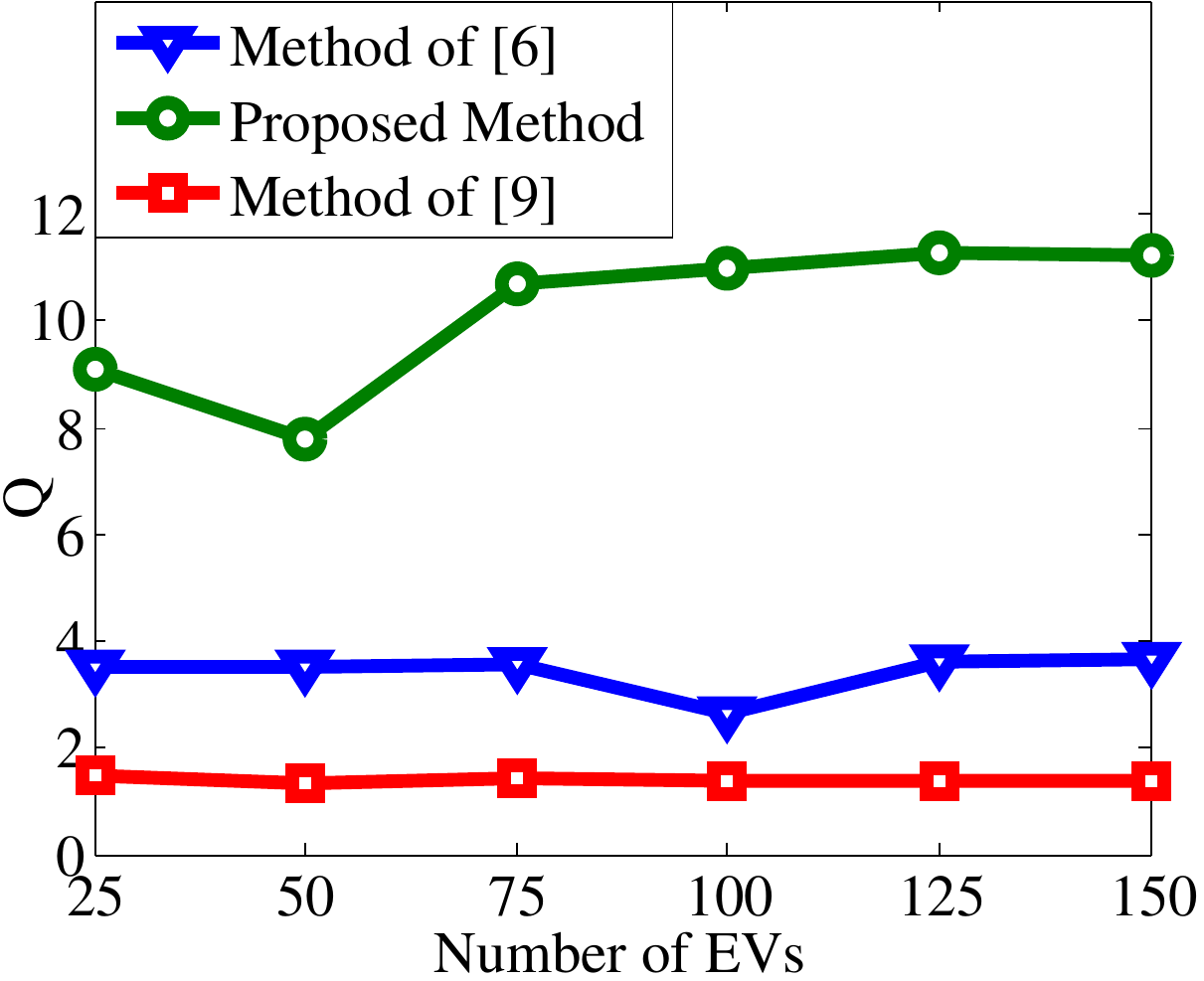}
				\caption{$\alpha=0.75$}
				\label{fig:alpha75}
			\end{center}
		\end{subfigure}
	\begin{subfigure}[b]{0.195\textwidth}
		\begin{center}
			\includegraphics[width=\textwidth]{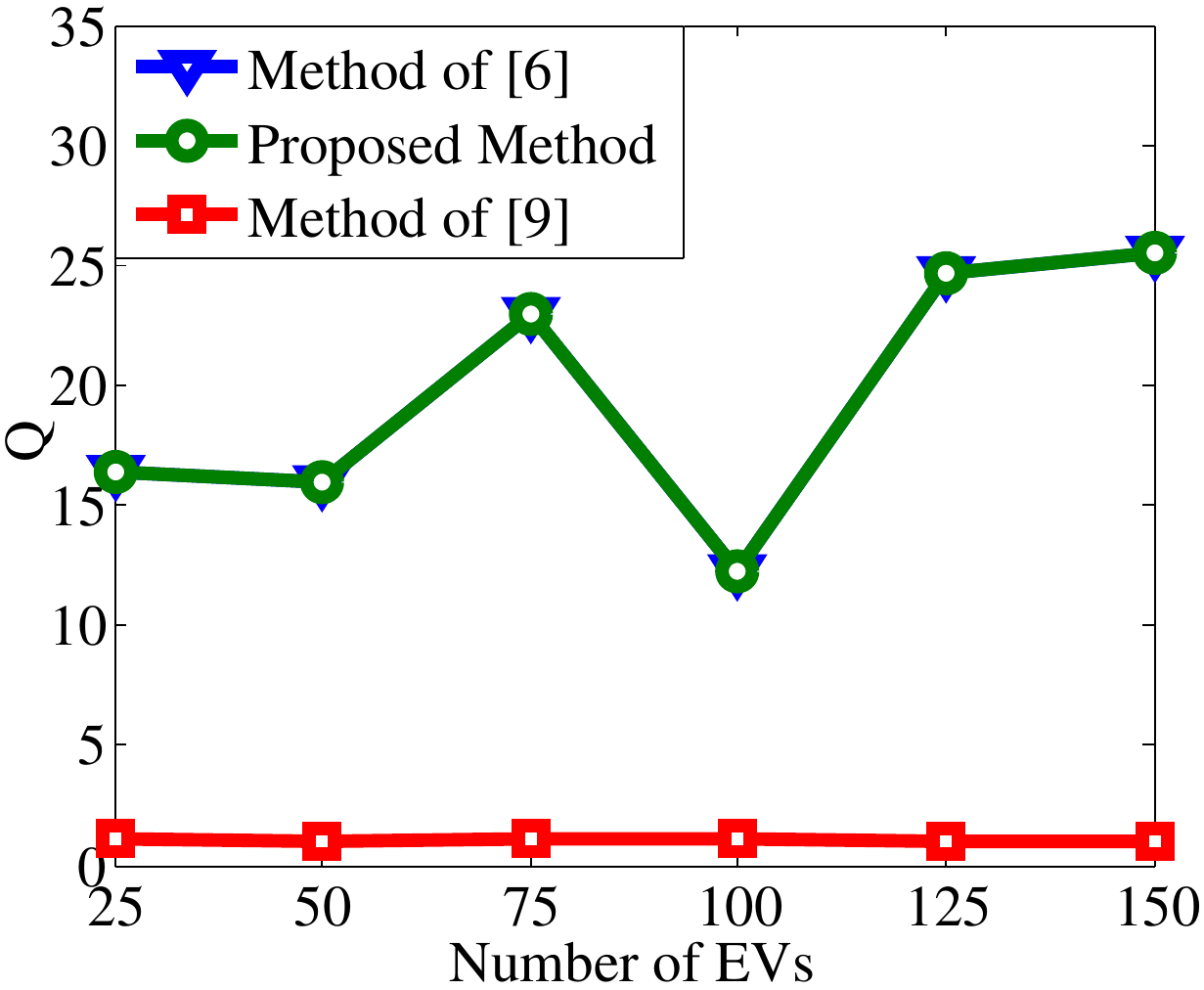}
			\caption{$\alpha=1$}
			\label{fig:alpha1}
		\end{center}
	\end{subfigure}%

	\caption{Comparison based on the defined performance criteria, $Q$ by varying $\alpha$ in Eq. (\ref{eq:Q}).} 
	\label{fig:Q-alpha}
\end{figure*}

We use Eq. (\ref{eq:Q}) as the underlying merit factor to compare three methods. Since the value of $\alpha$ is important in judging the result of the algorithms, we vary $\alpha$ from 0 to 1 to make the comparison fair. Fig. \ref{fig:Q-alpha} depicts the result for different values of $\alpha$ when the number of EVs change from 25 to 150. In Fig. \ref{fig:alpha0}, $\alpha$ is set to 0. In this case, $Q=J_2$ and only the value of user convenience is important. The user convenience maximization algorithm proposed in \cite{2012-wen-EV-selection} is better than the other two methods. The poor performance of method of \cite{2012-He-EV-selection} is because the algorithm is a cost minimization scheme and does not take care of the user convenience. In Fig. \ref{fig:alpha25}, $\alpha$ is increased to 0.25. The result is that our proposed algorithm provides better performance since both cost and user convenience are measured. When $\alpha =0.5$ in Fig. \ref{fig:alpha5}, both cost and user convenience have the same weight and the figure shows a better comparison of the three methods in terms of objective of problem $\mathcal{Z}$. Based on the depicted results, the proposed algorithm in this paper outperforms both Method of \cite{2012-He-EV-selection} and Method of \cite{2012-wen-EV-selection} significantly. This scenario implies that the proposed algorithm is better than the other methods, on average. For $\alpha =0.75$, Method of \cite{2012-He-EV-selection} is getting closer to our proposed algorithm since the weight of cost in evaluation is increased. Finally, by increasing the value of $\alpha$ to 1 in Fig. \ref{fig:alpha1}, the result for our proposed algorithm and Method of \cite{2012-He-EV-selection} become the same due to the fact that both obtain optimal charging cost. Method of \cite{2012-He-EV-selection} and the proposed algorithm show a substantial performance improvement to the Method of \cite{2012-wen-EV-selection}, in this case.

\subsection{Evaluation based on average charging time}
\begin{figure}
\begin{center}
\resizebox{1.8in}{!}{%
\includegraphics*{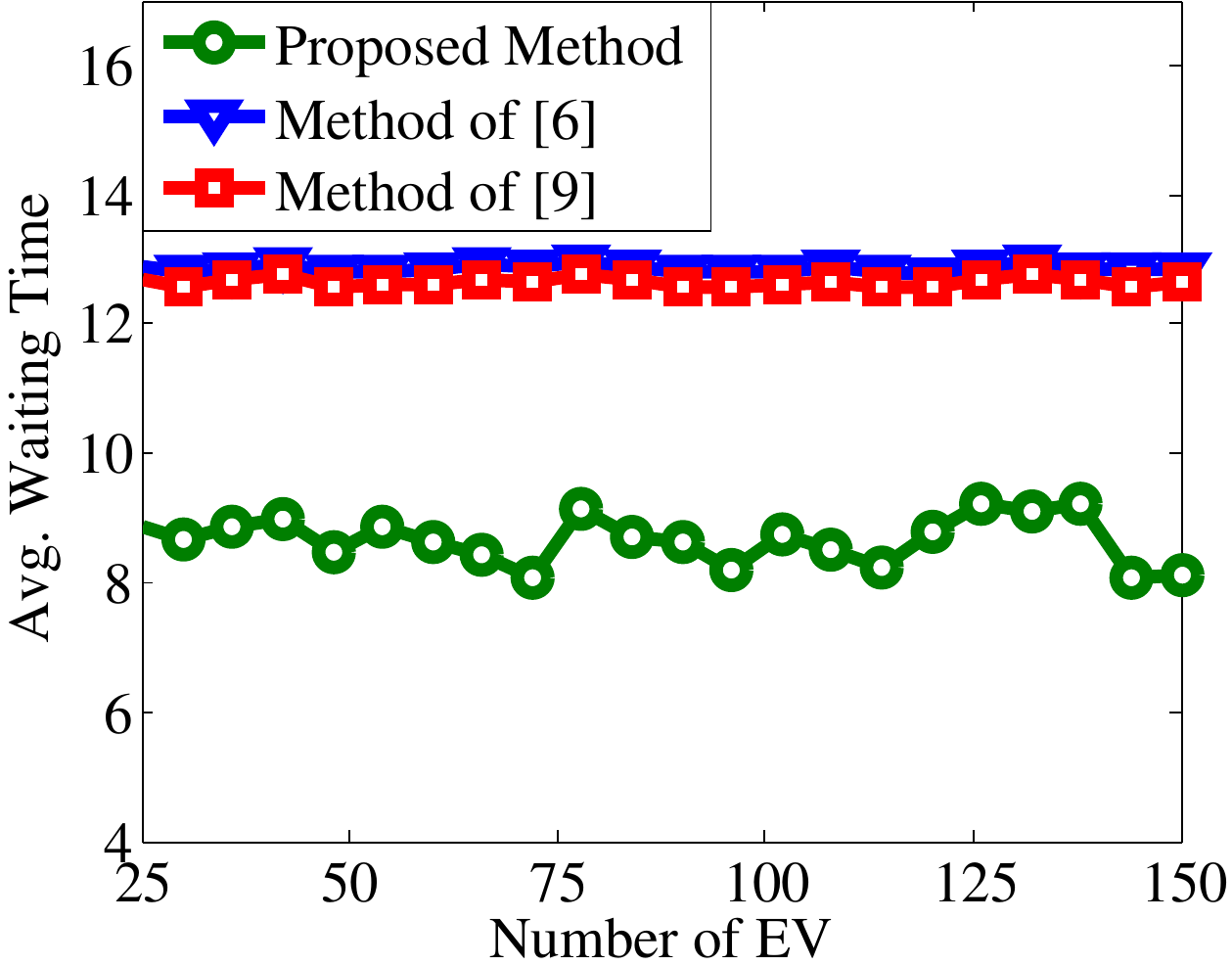} }%
\caption{Average charging time vs. Number of EVs}\label{fig:25resi_watiting_time_compare}
\end{center}
\end{figure}

The three described algorithms are compared in terms of average charging time when the total number of EVs change from 25 to 150 with step size 25. The waiting time is defined as the difference between arrival time and the finishing time of charging for each EV. Based on Fig. \ref{fig:25resi_watiting_time_compare}, the proposed method in this paper significantly improves the other methods by 36\%. We note that the main reason behind this is the policy of CSA algorithm in applying EDF strategy and charging EVs with their maximum charging rate as explained in Section \ref{sec:algorithm}. Method of \cite{2012-He-EV-selection} and Method of \cite{2012-wen-EV-selection} are very close to each other in terms of the average waiting time since non of these methods has a specific strategy to reduce waiting time and hence they act like a random method.

\section{Conclusion}\label{sec:conclusion}
This paper studies EV charging scheduling problem and unlike the previous studies which only focus on charging cost minimization or user convenience maximization, takes into account both factors and propose an efficient scheduling mechanism to the formulated bi-objective optimization problem as solution. The proposed algorithm is centralized and leverages load forecasting in order to respect power constraints imposed by the grid utility provider. The simulation results show the superiority of the scheduling mechanism.

As the future work, we are going to extend this study in two directions by proposing a distributed solution in one hand, and investigating a more realistic and mature definition of user convenience, on the other hand.

\newpage

\bibliography{IEEEabrv,references}

\end{document}